\documentclass[12pt,a4paper]{article}
\usepackage{comment}
\usepackage{amsmath}
\usepackage{amssymb}
\usepackage{amsthm}
\usepackage{amscd}
\usepackage{graphicx}
\usepackage{indentfirst}
\usepackage[ruled,vlined,linesnumbered]{algorithm2e}
\usepackage{titlesec}
\usepackage{titlesec}
\usepackage{color}
   \usepackage{amsmath,amssymb}
   \usepackage{latexsym}

\begin{document}
\newtheorem{definition}{\bf Definition}
\newtheorem{theorem}{\bf Theorem}
\newtheorem{proposition}{\bf Proposition}
\newtheorem{lemma}{\bf Lemma}
\newtheorem{corollary}{\bf Corollary}
\newtheorem{example}{\bf Example}
\newtheorem{remark}{\bf Remark}
\newtheorem{Table}{\bf Table}
\newtheorem{Sentence}{\bf Step}
\newtheorem{Branch}{}

%%%%%%%%%%%%%%%%%%%%%%%%%%%%%%%%%%%%%%%%%%%%%
%      Letters
%%%%%%%%%%%%%%%%%%%%%%%%%%%%%%%%%%%%%%%%%%%%%
\def\RRC {{\tt RealRootClassification}}
\def\RC {{\tt RegularChains}}
\def\T {\ensuremath{\bf{T}}}
\def\A {\ensuremath{\bf{A}}}
\def\B {\ensuremath{\bf{B}}}
\def\P {\ensuremath{\bf{P}}}
\def\S {\ensuremath{\mathbb{S}}}
\def\E {\ensuremath{\bf{E}}}
\def\H {\ensuremath{\bf{H}}}
\def\PF {\ensuremath{\bf {PF}}}
\def\TH {\ensuremath{\bf{TH}}}
\def\RS {\ensuremath{\mathbb{TH}}}
\def\HCTD {\ensuremath{\tt{HCTD}}}
\def\HPCTD {\ensuremath{\mathrm{HPCTD}}}
\def\CTD {\ensuremath{\mathrm{CTD}}}
\def\PCTD {\ensuremath{\mathrm{PCTD}}}
\def\SUS {\ensuremath{\mathrm{SUS}}}
\def\FUS {\ensuremath{\mathrm{FUS}}}
\def\WUCTD {\ensuremath{\mathrm{WUCTD}}}
\def\RSD {\ensuremath{\mathrm{RSD}}}
\def\FWCTD {\ensuremath{\mathrm{FWCTD}}}
\def\SWCTD {\ensuremath{\mathrm{SWCTD}}}
\def\ARSD {\ensuremath{\tt{RSD}}}
\def\APRSD {\ensuremath{\tt{PRSD}}}
\def\CCTD {\ensuremath{\tt{CTD}}}
\def\WUCTD {\ensuremath{\tt{WUCTD}}}
\def\ASUS {\ensuremath{\tt{SUS}}}
\def\AFUS {\ensuremath{\tt{FUS}}}
\def\ASWCTD {\ensuremath{\tt{SWCTD}}}
\def\ASMPD {\ensuremath{\tt{SMPD}}}
\def\AHPCTD {\ensuremath{\tt{HPCTD}}}
\newtheorem{Rules}{\bf Rule}

\title{ {Deciding Nonnegativity of Polynomials by MAPLE}}
\author{Lu Yang\\
         {\small CICA, Chinese Academy of Sciences, Chengdu, 610041}\\
         {\small Email: luyang@casit.ac.cn}\\
        Bican Xia\\
         {\small School of Mathematical Sciences}\\
         {\small Peking University, Beijing, 100871}\\
         {\small Email: xbc@math.pku.edu.cn}
         }
\date{}
\maketitle
\begin{abstract}
There have been some effective tools for solving (constant/parametric) semi-algebraic systems in Maple's library {\tt RegularChains} since Maple 13. By using the functions of the library, %especially the function
{\it e.g.},
\RRC, one can prove and discover polynomial inequalities. This paper is more or less a user guide on using \RRC\ to prove the nonnegativity of polynomials. We show by examples how to use this powerful
tool to prove a polynomial is nonnegative under some polynomial inequality and/or equation constraints. Some tricks for using the tool are also provided.

{\bf Keywords: } Automated reasoning, nonnegativity, Maple, RegularChains, RealRootClassification.
\end{abstract}

\section{Introduction}

Let us begin with an example.

\begin{example}
Prove that $f\ge 0$ under the constraints that $a\ge 0, b\ge 0, c\ge 0, abc-1=0,$ where
\[\begin{array}{rl}
f = & 2b^4c^4+2b^3c^4a+2b^4c^3a+2b^3c^3a^2+2a^3c^3b^2+2a^4c^3b+2a^3c^4b+2a^4c^4\\
& +2a^3b^4c+2a^4b^4+2a^3b^3c^2+2a^4b^3c-3b^5c^4a^3-6b^4c^4a^4-3b^5c^3a^4\\
& -3b^4c^3a^5-3b^4c^5a^3-3b^3c^5a^4-3b^3c^4a^5.
\end{array}
\]

To prove the inequality by Maple, we first start Maple and load two relative packages of \RC\ as follows.

\noindent\verb|> with(RegularChains):|\\
\verb|> with(ParametricSystemTools):|\\
\verb|> with(SemiAlgebraicSetTools):|

Then define an order of the unknowns:

\noindent\verb|> R := PolynomialRing([a, b, c]);|

Now, by calling

\noindent\verb|> RealRootClassification([abc-1], [a, b, c], [-f], [ ], 2, 0, R);|\\
we will know at once that the inequality holds.
\end{example}

In this paper, we give in detail an introduction on how to use the function \RRC\ ({\tt RRC} for short) to prove a polynomial is nonnegative under some polynomial inequality and/or equation constraints. Before we start, we would like to give some history remarks here.

It is well-known that Tarski \cite{tarski} proved that all elementary
algebraic and geometric propositions are decidable and gave an
algorithm for deciding whether or not a given elementary algebraic
and geometric proposition is true. Although Tarski's method cannot be applied to any non-trivial theorem
proving due to its high complexity, it is a milestone since, for the first time, it told us quantifier elimination (QE) in real closed fields is decidable. Collins \cite{CAD} proposed a so-called Cylindrical
Algebraic Decomposition (CAD) method in 1975. Although the CAD
method is of doubly exponential complexity, it has been successfully
applied to many non-trivial theorem proving and discovering. There
are many subsequent work which improved the CAD algorithm and have
been implemented as several well-known tools for solving general
QE problems, {\it e.g.}, {\tt QEPCAD}. %Nevertheless, for many problems such as the problems in this paper, the general methods often cannot get results.

%Wu \cite{wu77} proposed an efficient method in 1977, now called Wu's
%method, for proving theorems in plane geometry. Wu's method has
%gained a great success and a lot of subsequent work have been
%inspired. Wu's method concentrates on equality-type theorem proving
%\cite{wu84,zxq} and is not so successful for inequality-type theorem
%proving.

Yang {\it et. al.} \cite{yhz} gave a theorem for explicitly
determining the condition for a given polynomial to have a given
number of real (and/or complex) zeros. %The conditions are first order formulae with polynomials in the coefficients of the given polynomial.
Sometimes the conditions are called the
root-classification of the polynomial. With this theorem and its
generalization to the case of semi-algebraic system, Yang {\it et. al.}
proposed an algorithm for proving and discovering inequality-type theorems
automatically \cite{yhx, yang, xia}. Indeed, the algorithm solves a special kind of QE problems which have at least one polynomial equation. A key concept of the method is {\em border polynomial}. %The algorithm takes use of Wu's triangular decomposition method \cite{wu77,wu84} and CAD algorithm and
This algorithm has been improved
and implemented by Xia as a Maple package DISCOVERER
\cite{discover}. Since 2009, the main functions of DISCOVERER have
been integrated into the {\tt RegularChains} library of Maple.
Since then, the implementation has been improved by Chen {\it et. al.} \cite{chen12a,chen12b,chen13}. All the examples reported in this paper can be solved with Maple of version higher than Maple 13.

There are many other methods based on different principles for polynomial inequality proving. Since this paper is more or less a user guide on using \RRC\ to prove the nonnegativity of polynomials, we omit the introduction to those methods.

The rest of the paper is organized as follows. Section \ref{RC} describes the usage of the function
\RRC\ of {\tt RegularChains}. Section \ref{3} shows by examples how to use \RRC\ to prove a polynomial is nonnegative subject to some polynomial inequality and/or equation constraints. Some tricks for using the tool are also provided. %We conclude the paper in Section \ref{4}.

\section{\RRC}\label{RC}

In this section we describe in detail
the calling sequence, the input and output of {\tt
RealRootClassification} ({\tt RRC} for short).

First of all, you should install Maple in your computer. The
version of Maple should be at least Maple 13. Then, when Maple is
started, you should load the \RC\ library as follows before using
{\tt RRC}.

\noindent\verb|> with(RegularChains):|\\
\verb|> with(ParametricSystemTools):|\\
\verb|> with(SemiAlgebraicSetTools):|

The calling sequence of  {\tt RealRootClassification} is
\[{\tt RealRootClassification}(F, N, P, H, d, a, R);\]
where the first four parameter $F, N, P$ and $H$ represent a semi-algebraic system of the following form
\[F=0, N\geq 0, P>0, N\neq 0.\]
Herein, each of $F, N, P$ and $H$ is a set of polynomials in unknowns
$x_1,...,x_n$ with rational coefficients. If $F=[f_1,...,f_s],$
$N=[g_1,...,g_t],$ $P=[p_1,...,p_k],$ and $H=[h_1,...,h_m],$ then
$F=0, N\geq 0, P>0, N\neq 0$ is a short form for the following
system
\[\left\{\begin{array}{l}
f_1=0,...,f_s=0,\\
g_1\geq 0,...,g_t\geq 0,\\
p_1>0,...,p_k>0,\\
h_1\ne 0,...,h_m\ne 0.
\end{array}\right.\]

It should be pointed out that $s$ must be positive, {\it i.e.}, the system must have at least one equation.

The last formal parameter $R$ is a list of the variables
$x_1,...,x_n$, which defines an order of the variables and should be defined as a type {\em PolynomialRing} (see Example \ref{ex:1}).

The formal
parameter $d$ is a positive integer which indicates the last $d$
elements in $R$ are to be viewed as parameters of the given system.

The formal parameter $a$ has two possible forms. If $a$ is a
nonnegative integer, then \RRC\ will output the conditions for the
system $[F=0, N\geq 0, P>0, N\neq 0]$ to have exactly $a$ distinct
real solutions. If $a$ is a range, {\it e.g.} $2..3$, then \RRC\ will
output the conditions for the number of distinct real solutions of
the system $[F=0, N\geq 0, P>0, N\neq 0]$ falls into the range $a$.
If the second element of a range is an unassigned name, it means positive infinity.

We illustrate the usage of \RRC\ by the following simple example.

\begin{example}\label{ex:1}
We want to know the conditions on the coefficients of $f=ax^2+bx+c$
for $f$ to have real roots if $a\ne 0.$
\end{example}

After loading \RC\ library and two relative packages, we define the system as follows.

\noindent\verb|> f:=a*x^2+b*x+c;|\\
\verb|> F:=[f]; N:=[ ]; P:=[ ]; H:=[a];|\\
\verb|> R:=PolynomialRing([x,a,b,c]);|

To get more information from the output of the function directly, we
type in:\\
\verb|> infolevel[RegularChains]:=1;|

Then, we call\\
\verb|> RealRootClassification(F, N, P, H, 3, 1..n, R);|\\
where the range $1..n$ means ``the polynomial has at
least one real roots".

The output is: $R_1>0$ where $R_1=b^2-4ac$ provided that $a\ne 0$
and $R_1\ne 0.$ To discuss the case when $R_1=0,$ we can add this
equation into the original system and call \RRC\ again.\\
\begin{verb}
> RealRootClassification([b^2-4*a*c,op(F)], N, P, H, 3, 1..n,
R);
\end{verb}

In this way, we finally know that the condition is $R_1\geq 0.$

\section{Deciding nonnegativity by {\tt RRC}}\label{3}

We first give a detailed explanation of Example 1.

\noindent{\bf Example 1} (continued).\ Obviously,
\[a\ge 0 \wedge b\ge 0 \wedge c\ge 0 \wedge abc-1=0 \Longrightarrow f\ge 0 \]
is equivalent to the following system is inconsistent
\[a\ge 0 \wedge b\ge 0 \wedge c\ge 0 \wedge abc-1=0 \wedge f<0.\]
So, in Example 1, we call

\noindent\verb|> RealRootClassification([abc-1], [a, b, c], [-f], [ ], 2, 0, R);|\\
where the ``0" means we want to compute the conditions for the system to have no real solutions.

The output is:
\begin{center}
There is always given number of real solution(s)!\\
PROVIDED THAT\\
$\phi(b,c)\ne 0,$
\end{center}
where $\phi(b,c)$ is a polynomial in $b$ and $c$ with $19$ terms and of degree $18.$

The output means that the system always has no real solutions provided that the polynomial $\phi(b,c)$ does not vanish. In other word, {\tt RRC} proves that the proposition holds for almost all $a,b$ and $c$ except those such that $\phi(b,c)=0$.

Because the inequality to be proved is a non-strict inequality ($f\ge 0$), by continuity, we know at once that $f\ge 0$ holds for all $a,b$ and $c$ such that $a\ge 0 \wedge b\ge 0 \wedge c\ge 0 \wedge abc-1=0$. Thus, the proposition is proved.

\begin{example}
Prove that \[a\ge 0 \wedge b\ge 0 \wedge c\ge 0 \wedge ab+bc+ca-1=0 \Longrightarrow g\ge 0\]
where
\[\begin{array}{rl}
g = & -10a^3b^3-10b^3c^3-10a^3c^3-5a^4b^2-5c^2a^4-5c^4a^2-5a^2b^4+4c^3a\\
 & -5b^4c^2-5b^2c^4+4ca^3+2a^4+2b^4+2c^4-10cab^4-30c^2a^3b-10ca^4b\\
 & -10c^4ab+4a^3b^4c+16a^3b^3c^2+4a^4b^3c+16b^3c^3a^2+16a^3c^3b^2\\
 & +4a^3c^4b+4b^3c^4a+4b^4c^3a+4a^4c^3b+6b^2c^2a^4-30b^3c^2a\\
 & -30c^3a^2b+6b^2c^4a^2+16c^2ab+16ca^2b-50b^2c^2a^2+16cab^2-30b^2c^3a\\
 & -30ca^3b^2-30a^2b^3c+6b^4c^2a^2+6c^2a^2+6a^2b^2+6b^2c^2+4c^3b+4b^3c\\
 & +4b^3a+4a^3b+2a^4b^4+2a^4c^4+2b^4c^4.
\end{array}\]
\end{example}

\begin{example}
Prove that \[x\ge 0 \wedge y\ge 0 \wedge z\ge 0 \wedge r \ge 0 \wedge (r+1)^2-4/3 \ge 0 \wedge x+y+z-3=0 \Longrightarrow h\ge 0\]
where
\[\begin{array}{rl}
h = & -3+z-3r^3y^2z^2x^2+ry^3+r^2z^3+rz^3-3ry^2-3rz^2+r^2x^3+yr\\
 & +r^2y^3+zr+rx^3+rx-3rx^2+xr^2z^2+yrx^2+xrz^2+r^3y^3x^2\\
 & +r^2y^3x^2-3r^2y^2z^2+r^2y^2z^3+r^3z^2x^3+r^2z^2x^3+zr^2y^2+zry^2\\
 & +yr^2x^2-3r^2z^2x^2-3r^2y^2x^2+r^3y^2z^3+y+x.
\end{array}\]

\end{example}

\begin{example}
Prove that \[a\ge 0 \wedge b\ge 0 \wedge c\ge 0 \wedge d \ge 0\wedge a+b+c+d-1=0 \Longrightarrow p\ge 0\]
where
\[p = 1+176abcd-27(bcd+cda+dab+abc).\]
\end{example}

\begin{example}
Prove that for any given integer $n\ge 3$,
\[-1\le x_i\le 1\ (1\le i\le n) \wedge \sum{x_i^3}=0 \Longrightarrow \sum{x_i}\le \frac{n}{3}.\]
Although the problem is not so hard for a mathematician, it is really hard for a computer. We proved the proposition for $n=3,4,5$ by Maple.
\end{example}

\begin{example}
Prove that \[a\ge 0 \wedge b\ge 0 \wedge c\ge 0 \wedge a^3b+b^3c+c^3a-3=0 \Longrightarrow q\ge 0\]
where
\[q = -75a^4b^4c^4-5a^4b^4-5a^4c^4-5b^4c^4+21a^4+21b^4+21c^4+27.\]
\end{example}

\begin{example}\footnote{http://www.artofproblemsolving.com/Forum/viewtopic.php?f=52\&t=432676}
Prove that \[a\ge 0 \wedge b\ge 0 \wedge c\ge 0 \wedge a^3b+ac^3+b^3c+abc-4=0 \Longrightarrow w\ge 0\]
where
\[w = 27(a+b+c)^4-1024.\]
\end{example}

Examples 3-8 have a common property that the systems themselves have at least one equation. So, we can use \RRC\ directly. We show by the following two examples how to deal with the situation where no equations appear in the system.

\begin{example}
Prove that \[a\ge 0 \wedge b\ge 0 \wedge c\ge 0 \wedge d\ge 0 \Longrightarrow u\ge 0\]
where
\[\begin{array}{rl}
u = & 1280bd^3c+624bc^2d^2+320ab^4+464ac^4-112ad^4-112a^4b+464a^4c\\
 & -112b^4c+464b^4d+208c^3b^2+1072d^3b^2-224b^3c^2+1072b^3d^2\\
 & +320bc^4+464bd^4-112c^4d+208d^3c^2-224c^3d^2+320cd^4+128ad^3c\\
 & +624ab^2c^2+740b^3cd+1812ab^2d^2+516ac^2d^2+1812b^2cd^2\\
 & +128bc^3d+516b^2c^2d+128a^3bd+624a^2b^2d+516a^2bd^2+1280a^3cd\\
 & +1812a^2c^2d+624a^2cd^2+128ab^3c+1280ab^3d+1280ac^3b+740ac^3d\\
 & +740ad^3b+1812a^2bc^2+740a^3bc+516a^2b^2c+1896ab^2cd+1896abc^2d\\
 & +1896abcd^2+1896a^2bcd+320a^4d+208b^3a^2+1072c^3a^2-224d^3a^2\\
 & -224a^3b^2+1072a^3c^2+208a^3d^2+64a^5+64b^5+64c^5+64d^5.
\end{array}\]
As usual, we want to prove that the following system has no real solutions
\[a\ge 0 \wedge b\ge 0 \wedge c\ge 0 \wedge d\ge 0 \wedge u<0.\]
However, the system does not contain equations and thus {\tt RRC} cannot be applied directly.

We introduce a new variable $T$ and the system being inconsistent is equivalent to that the following new system is inconsistent
\[a\ge 0 \wedge b\ge 0 \wedge c\ge 0 \wedge d\ge 0 \wedge u+T=0 \wedge T>0.\]
For this new problem, we first define

\noindent\verb|> R := PolynomialRing([T, a, b, c, d]);|\\
and then call

\noindent\verb|> RealRootClassification([u+T], [a, b, c, d], [T], [ ], 4, 0, R);|\\
The problem is solved immediately.
\end{example}

\begin{example}
Prove that \[a\ge 0 \wedge b\ge 0 \wedge c\ge 0 \Longrightarrow v\ge 0\]
where {\small
\[\begin{array}{rl}
v = & 104976a^{12}+1679616a^{11}b+1469664a^{11}c+10850112a^{10}b^2\\
 & +19046016a^{10}bc+8076024a^{10}c^2+36149760a^9b^3+95364864a^9b^2c\\
 & +80561952a^9bc^2+22935528a^9c^3+65762656a^8b^4+228601856a^8b^3c\\
 & +282635520a^8b^2c^2+162625040a^8bc^3+42710593a^8c^4+63474176a^7b^5\\
 & +251921856a^7b^4c+354740704a^7b^3c^2+288770224a^7b^2c^3\\
 & +207550776a^7bc^4+83017484a^7c^5+29076288a^6b^6+60534016a^6b^5c\\
 & -155234320a^6b^4c^2-380047056a^6b^3c^3+3130676a^6b^2c^4\\
 & +375984436a^6bc^5+181119606a^6c^6+8313344a^5b^7-89738240a^5b^6c\\
 & -760459488a^5b^5c^2-1768157568a^5b^4c^3-1403613720a^5b^3c^4\\
 & +236428572a^5b^2c^5+824797636a^5bc^6+291288188a^5c^7\\
 & +13943056a^4b^8-3628032a^4b^7c-514131904a^4b^6c^2-1869896304a^4b^5c^3\\
 & -2495402586a^4b^4c^4-783163260a^4b^3c^5+1171287578a^4b^2c^6\\
 & +1122586500a^4bc^7+288706561a^4c^8+18028800a^3b^9+116005472a^3b^8c\\
 & +171678496a^3b^7c^2-347011440a^3b^6c^3-1231272792a^3b^5c^4\\
 & -894635820a^3b^4c^5+731754984a^3b^3c^6+1497257080a^3b^2c^7\\
 & +851454308a^3bc^8+170469720a^3c^9+10593792a^2b^{10}+100409472a^2b^9c\\
 & +365510616a^2b^8c^2+624203728a^2b^7c^3+480156788a^2b^6c^4\\
 & +215762988a^2b^5c^5+511667522a^2b^4c^6+990571720a^2b^3c^7\\
 & +861820134a^2b^2c^8+356931720a^2bc^9+58375800a^2c^{10}\\
 & +2985984ab^{11}+34730496ab^{10}c+165207744ab^9c^2+415788248ab^8c^3\\
 & +606389880ab^7c^4+560561092ab^6c^5+437187748ab^5c^6+422470380ab^4c^7\\
 & +390424292ab^3c^8+235263240ab^2c^9+77497200abc^{10}+10692000ac^{11}\\
 & +331776b^{12}+4478976b^{11}c+25292160b^{10}c^2+77899104b^9c^3\\
 & +144247489b^8c^4+170606684b^7c^5+141892350b^6c^6+102086036b^5c^7\\
 & +76748161b^4c^8+52182360b^3c^9+24766200b^2c^{10}+6804000bc^{11}\\
 & +810000c^{12}.
\end{array}\] }
Similar to Example 8, the inequality is proved by first defining

\noindent\verb|> R := PolynomialRing([T, a, b, c]);|\\
and then calling

\noindent\verb|> RealRootClassification([v+T], [a, b, c], [T], [ ], 3, 0, R);|
\end{example}

We report the timings on the examples in the following table.
All the computation were performed on a computer (CPU 3.2GHz, 2G RAM, Windows XP) with Maple 13.

%\begin{Table}\label{t1}
\begin{center}
\begin{tabular}{|r|r|r|}
     \hline
     No.  & timing   & memory\\
     \hline
     {\em EX1} &0.06s&   0.81M\\
     {\em EX3} &0.04s&   0.81M\\
     {\em EX4} &6.04s&  53.55M\\
     {\em EX5} &0.03s&   0.81M\\
     {\em EX6(n=5)} &377.35s& 118.60M\\
     {\em EX7} &16.67s&  63.11M\\
     {\em EX8} &2.98s&  44.67M\\
     {\em EX9} &1.26s&  39.36M\\
     {\em EX10}&0.57s&  38.86M\\
     \hline
\end{tabular}
\end{center}
%\end{Table}

%\section{Conclusion}\label{4}

\end{document}